# A Detailed Survey on Various Aspects of SQL Injection in Web Applications: Vulnerabilities, Innovative Attacks, and Remedies

Diallo Abdoulaye Kindy[1,2] and Al-Sakib Khan Pathan[2]

[1]CustomWare, Kuala Lumpur, Malaysia
[2]Department of Computer Science, International Islamic University Malaysia, Kuala Lumpur, Malaysia
diallo14@gmail.com and sakib@iium.edu.my

*Abstract*: In today's world, Web applications play a very important role in individual life as well as in any country's development. Web applications have gone through a very rapid growth in the recent years and their adoption is moving faster than that was expected few years ago. Now-a-days, billions of transactions are done online with the aid of different Web applications. Though these applications are used by hundreds of people, in many cases the security level is weak, which makes them vulnerable to get compromised. In most of the scenarios, a user has to be identified before any communication is established with the backend database. An arbitrary user should not be allowed access to the system without proof of valid credentials. However, a crafted injection gives access to unauthorized users. This is mostly accomplished via SQL Injection input. In spite of the development of different approaches to prevent SQL injection, it still remains an alarming threat to Web applications. In this paper, we present a detailed survey on various types of SQL Injection vulnerabilities, attacks, and their prevention techniques. Alongside presenting our findings from the study, we also note down future expectations and possible development of countermeasures against SQL Injection attacks.

## 1. Introduction

In the recent years, the World Wide Web (WWW) has witnessed a staggering growth of many online Web applications which have been developed for meeting various purposes. Now-a-days, almost everyone in touch with '*computer technology*' is somehow connected online. To serve this huge number of users, great volumes of data are stored in Web application databases in different parts of the globe. From time to time, the users need to interact with the backend databases via the user interfaces for various tasks such as: updating data, making queries, extracting data, and so forth. For all these operations, design interface plays a crucial role, the quality of which has a great impact on the security of the stored data in the database. A less secure Web application design may allow crafted injection and malicious update on the backend database. This trend can cause lots of damages and thefts of trusted users' sensitive data by unauthorized users. In the worst case, the attacker may gain full control over the Web application and totally destroy or damage the system. This is successfully achieved, in general, via SQL injection attacks on the online Web application database. In this paper, we have reviewed most of the well-known and new SQL Injection attacks, vulnerabilities and prevention techniques. We present this topic in a way that the work could be beneficial both for the general readers and for the researchers in the area for their future research works.

SQL Injection is a type of injection or attack in a Web application, in which the attacker provides Structured Query Language (SQL) code to a user input box of a Web form to gain unauthorized and unlimited access. The attacker's input is transmitted into an SQL query in such a way that it forms an SQL code [1], [10]. In fact, SQL Injection is categorized as the top-10 2010 Web application vulnerabilities experienced by Web applications according to OWASP (Open Web Application Security Project) [9].

SQL Injection Vulnerabilities (SQLIVs) are one of the open doors for hackers to explore. Hence, they constitute a severe threat for Web application contents. The key root and basis of SQLIVs is quite simple and well understood: insufficient validation of user input [1]. To mitigate these vulnerabilities, many prevention techniques have been suggested such as manual approach, automated approach; secure coding practices, static analysis, using prepared statements, and so forth. Though, proposed approaches have achieved their goals to some extent, SQL Injection Vulnerabilities in Web applications remain as a major concern among application developers.

Relating to the above mentioned texts, the key objective of this work is to present a detailed survey on various types of SQL Injection vulnerabilities, attacks, and their prevention techniques. Alongside presenting our findings from the study, we also note down future expectations and possible development of countermeasures against SQL Injection attacks. The key purpose of this study is to address the issue from all necessary angles so that the work could be used as a reference work by the researchers and practitioners. Till today, a comprehensive survey on this topic is missing; hence, we believe our work could fill the void.

Though there are some previous works on SQL Injections, they have mainly the following limitations:

- *Not up-to-date:* the growth of e-commerce is almost parallel to the alarming threats targeting Web applications using SQL Injections. Hence, the relevance and accuracy of some previous publications are now questionable. The more the time passes by, the more kinds of attacks evolve and put less confidence on the previously noted information.

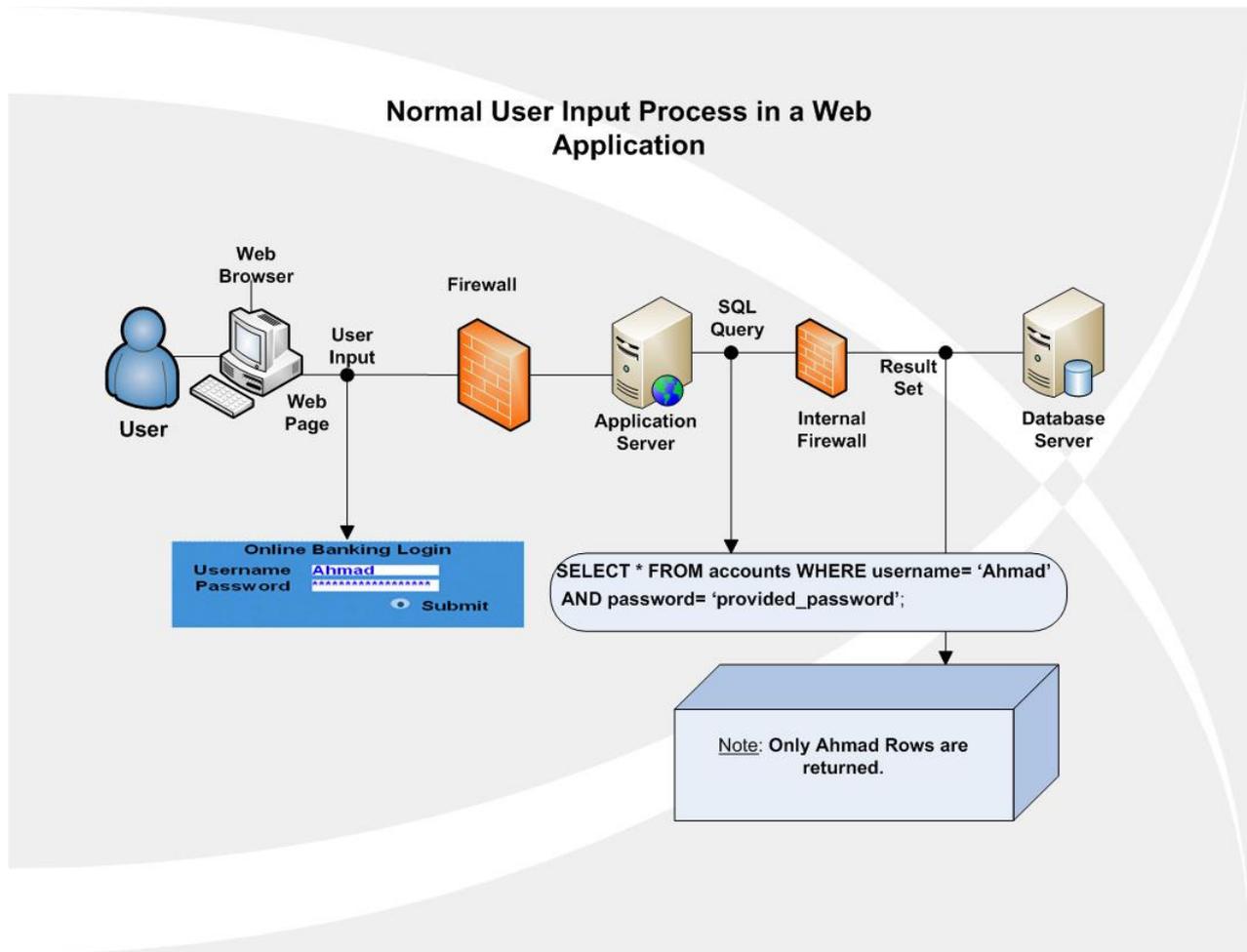

**Figure 1.** Normal user input process in a Web application.

- *Lack of practice:* In almost all the previous works, there is a critical lack of a discussion about the Web application security training tutorials used in practice. Sometimes, there is huge gap between theory and practice. Hence, in our work we mention the tools that should be known for practical use and tackling SQL injection attacks. The information about these tools is missing in most, if not all of the previous works we have analyzed.

After this initial information, the rest of the paper is organized as follows: in Section 2, we provide SQL Injection background and categorize the vulnerabilities and attacks. Section 3 presents an in-depth look at the most common SQL Injection attacks. Section 4 notes down the tools and tutorials that we have used for our work, Section 5 talks about various approaches for detecting SQL Injection attacks, Section 6 notes down the available countermeasures to tackle various SQL Injection attacks and a comparative analysis of various attacks and schemes, and finally, Section 7 concludes the paper noting the contribution of this work alongside mentioning our future research objectives.

## 2. SQL Injection: The 'Need-to-Know' Aspects

### 2.1 What is SQL?

SQL (pronounced as "S-Q-L" or "sequel") stands for Structured Query Language. It is the high level language used in various relational Database Management Systems (DBMS). SQL was originally developed in the early 1970's by Edgar F. Codd at IBM. It was commercial and the most-widely used language for all relational databases. This language is a declarative computer language which has elements that include clauses, expressions, predicates, queries, and statements. It allows the users mainly, (i) data insertion, (ii) data updating, (iii) query, (iv) deletion, and many more features (thus gives the user the power of manipulating databases) [6], [7].

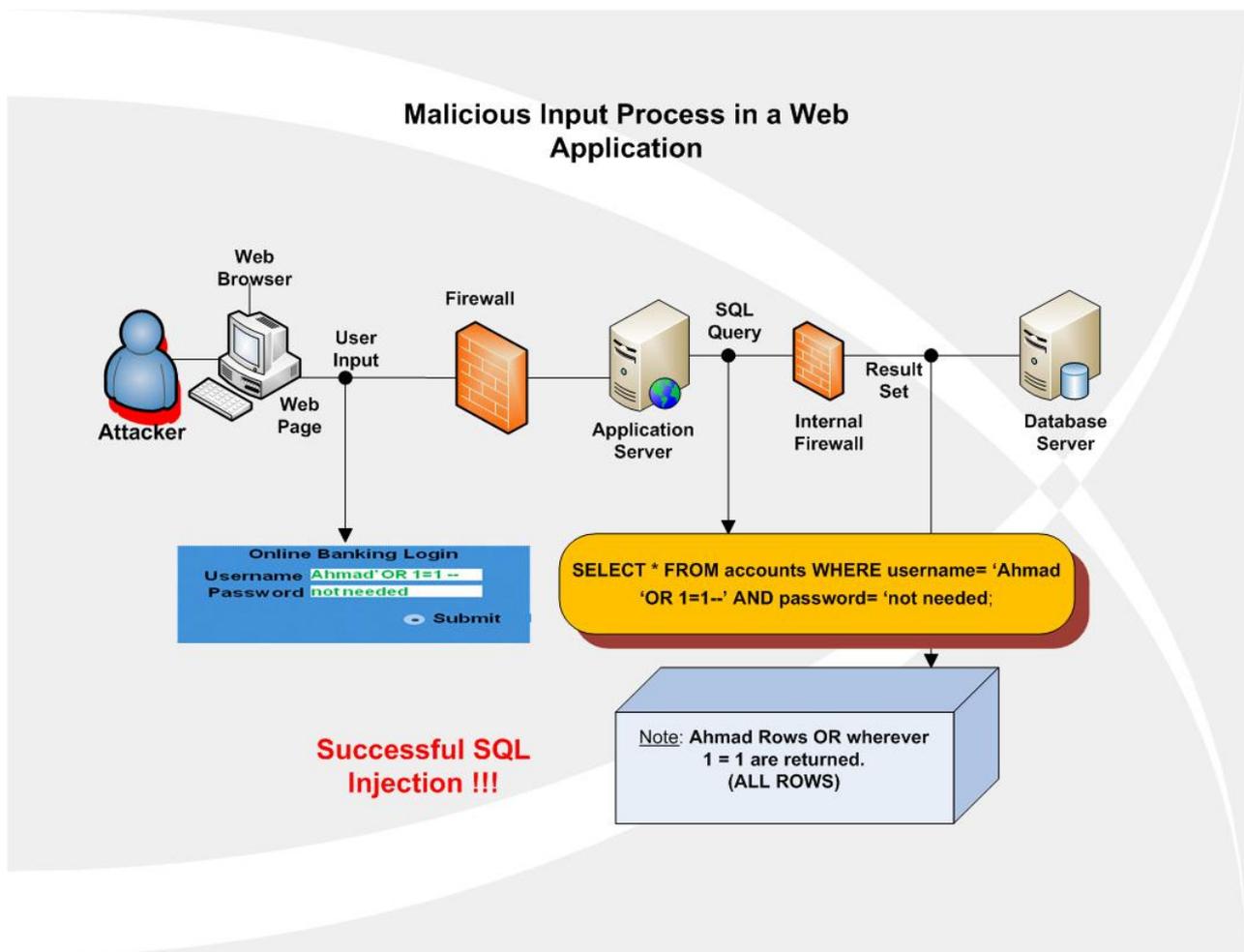

**Figure 2.** Malicious input process in a Web application.

### 2.2 SQL Injection Vulnerability versus SQL Injection Attack

Vulnerability in any system is defined as a bug, loophole, weakness or flaw existing in the system that can be exploited by an unauthorized user in order to gain unlimited access to the stored data. Attack generally means an illegal access, gained through well crafted mechanisms, to an application or system. An SQL Injection Attack (SQLIA) is a type of attack [30] whereby an attacker (a crafted user) adds malicious keywords or operators into an SQL query (e.g., SQL malicious code statements), then injects it to a user input box of a Web application. This allows the attacker to have illegal and unrestricted access to the data stored at the backend database. Figure 1 shows the normal user input process in a Web application, which is self-explanatory. Figure 2 shows an example how a malicious input could be processed in a Web application. In this case, the malicious input is the carefully formulated SQL query which passes through the system's verification method. To explore this area more, in this paper, we investigate both the SQL Injection vulnerabilities and SQL Injection attacks.

### 2.3 Why is SQL Injection a Threat?

Injecting a Web application is the synonym of having illegal access to the data stored in the database. The data sometimes could be confidential and of high value like the financial secrets of a bank or list of financial transactions or secret information of some kind of information system. An unauthorized access to this data by a crafted user can pose threat to their confidentiality, integrity, and authority. As a result, the system could bear heavy loss in giving proper services to its users or it may even face complete destruction. Sometimes such type of collapse of a system can threaten the existence of a company or a bank or an industry. If it happens against the information system of a hospital, the private information of the patients may be leaked out which could threaten their reputation or may become a case of defamation. Attackers may even use such type of attack to get confidential information that is related to the national security of a country. Hence, SQL Injection could be very dangerous in many cases depending on the platform where the attack is launched and where it gets success in injecting rogue users to the target system.

### 2.4 Types of Vulnerabilities in Web Programming Languages

There could be various types of vulnerabilities that could be exploited for SQL Injection. In this section, we present the most common security vulnerabilities found in Web programming languages [12] through which SQL Injection attacks are usually launched. We show the major types of vulnerabilities at a glance in Table 1.

**Table 1.** Types of Vulnerabilities at a glance

| Vulnerability Types | Basic Idea |
|---|---|
| *Type I* | Lack of clear distinction between data types accepted as input in the programming language used for the Web application development. |
| *Type II* | Delay of operation analysis till the runtime phase where the current variables are considered rather than the source code expressions. |
| *Type III* | Weak concern of type specification in the design: a number can be used as a string or vice-versa. |
| *Type IV* | The validation of the user input is not well defined or sanitized. Inputs are not checked correctly. |

### 2.5 Types of SQL Injection Attacks (SQLIAs): Past and Present

It is not an easy task to find out and categorize all types of SQLIAs. The same attack may be called with different names in different cases depending on the system scenario. In this sub-section, we present all the commonly known SQL Injection attacks [1], [11] that so far have been discovered along with newly invented innovative attacks. We use the terminologies as deem to be appropriate. Table 2 shows the types of SQL Injection attacks with brief descriptions.

## 3. An In-Depth Look At the Most common SQL Injection Attacks

Among various types of SQLI attacks, some are frequently used by the attackers. It is imperative to know the commonly used major attacks among all available attacks. Hence, in this section, we present an in-depth look at some of the most common SQL Injection attacks. We explain each of these major attacks with simple examples, wherever appropriate.

### 3.1 Tautology

SQL injection codes are injected into one or more conditional statements so that they are always evaluated to be true. Under this technique, we may have the following types and scenarios of attacks:

#### 3.1.1 String SQL Injection

This type of injection is also referred to as AND/OR Attack [14], [15]. The attacker inputs SQL tokens or strings to a conditional query statement that always evaluates to a true statement. The interesting issue with this type of attack is that instead of returning only one row in a table, if it is successful, it causes all of the rows in the database table targeted by the query to be returned. The goal behind this type of attack may include the following: (a) Bypassing authentication, (b) Identifying parameters that can be injected, and (c) Extraction of data [1].

**Scenario**

- **Normal Statement: SELECT * FROM users WHERE name='Lucia01**
  **Input:** Lucia01　**Output:** Lucia's Rows only
- **Injected Statement: SELECT * FROM users WHERE name= 'Lucia01' OR '1' ='1'**
  **Input:** 'Lucia01' OR '1' ='1'
  **Output:** this will return rows for Lucia01 OR wherever one equals to one which is true for all rows. Hence, all rows will be returned.

#### 3.1.2 Numeric SQL Injection

This type of Injection is quasi-similar to the above discussed. The main difference is that; here numeric values are used instead of strings. Therefore, the attacker would input numeric values to a conditional query statement that would always evaluate to a true statement.

**Scenario**

- **Normal Statement: SELECT * FROM users WHERE id= '101'**
  **Input:** 101 **Output:** id 101's Rows only.
- **Injected Statement: SELECT * FROM users WHERE name= '101' OR '1' ='1'.**
  **Input:** '101' OR '1' ='1'
  **Output:** this will return rows for '101'id or wherever one equals to one **(ALL ROWS)**
  **Note:** the crafted user can be more specific by adding ORDER BY clause to get exactly what he wants on time. The malicious input will look like: 101 OR 1=1 ORDER BY salary desc;

#### 3.1.3 Comments Attack

This type of attack takes advantage of the inline commenting allowed by SQL [29] - the malicious code and comments whatever comes after the "--" in the WHERE clause. The point is that everything after the comment characters will be ignored. Comments Attack can be combined with either String or Numeric SQL Injection so that it performs as a tautology which always evaluates to a true statement.

**Scenario**

- User Input: 'user1 OR '1' ='1 —'.
- Generated SQL Query: **SELECT username, password FROM clients WHERE username = 'user1 OR '1' ='1 —' AND password = 'whatever'.**

In this case, not only the WHERE clause is transformed into a tautology by the (OR 1=1) but also the password part is also completely ignored, hence only the username part will be checked [1], [29].

### 3.2 Inference

An attacker derives logical conclusions from the answer to a true/false question concerning the database. Through a successful inference, crafted users change the behavior of the database.

**Table 2.** Types of SQLIAs at a glance.

| Types of Attack | Working Method |
|---|---|
| *Tautologies* | SQL injection codes are injected into one or more conditional statements so that they are always evaluated to be true. |
| *Logically Incorrect Queries* | Using error messages rejected by the database to find useful data facilitating injection of the backend database. |
| *Union Query* | Injected query is joined with a safe query using the keyword UNION in order to get information related to other tables from the application. |
| *Stored Procedure* | Many databases have built-in stored procedures. The attacker executes these built-in functions using malicious SQL Injection codes. |
| *Piggy-Backed Queries* | Additional malicious queries are inserted into an original injected query. |
| *Inference*<br>- *Blind Injection*<br>- *Timing Attacks* | An attacker derives logical conclusions from the answer to a true/false question concerning the database.<br>- Information is collected by inferring from the replies of the page after questioning the server true/false questions.<br>- An attacker collects information by observing the response time (behavior) of the database. |
| *Alternate Encodings* | It aims to avoid being identified by secure defensive coding and automated prevention mechanisms. It is usually combined with other attack techniques. |

*3.2.1 Blind SQL Injection*

In this type of attack, useful information for exploiting the backend database is collected by inferring from the replies of the page after questioning the server some true/false questions. It is very similar to a normal SQL Injection [14], [15]. However, when an attacker attempts to exploit an application, rather than getting a useful error message, they get a generic page specified by the developer instead. This makes exploiting a potential SQL Injection attack more difficult but not impossible. An attacker can still get access to sensitive data by asking a series of True and False questions through SQL statements.

**Scenario**

```
http://victim/listproducts.asp?cat=boo
ks
SELECT * from PRODUCTS WHERE
category='books'
http://victim/listproducts.asp?cat=boo
ks' or '1'='1.
SELECT * from PRODUCTS WHERE
category='books' or '1'='1'.
```

*3.2.2 Timing Attacks*

An attacker collects information by observing the response time (behavior) of the database. Here the main concern is to observe the response time that will help the attacker to decide wisely on the appropriate injection approach.

*3.2.3 Database Backdoors*

Databases are used not only for data storage but also to keep malicious activity like a trigger. In this case, an attacker can set a trigger in order to get the user input and get it directed to his or her e-mail, for example.

**Scenario**

```
101; CREATE TRIGGER myBackDoor BEFORE
INSERT ON employee FOR EACH ROW BEGIN
UPDATE employee SET
email='hacker@me.com'WHERE userid =
NEW.userid.
```

*3.2.4 Command SQL Injection*

The purpose of this injection is to inject and execute commands specified by the hacker in the vulnerable application. The application executing the unwanted system commands is like a pseudo system-shell controlled by the attacker. Lack of correct input data validation (forms, cookies, HTTP headers, etc.) is the main vulnerability exploited by attackers for a successful injection. It differs from code injection in the sense that the attacker adds his own code to the existing code. Hence, the default functionalities of the application are extended without executing system commands. An OS (Operating System) command injection attack occurs when an attacker attempts to execute system level commands through a vulnerable application. Applications are considered vulnerable to OS command injection attack if they utilize user input in a system level command.

**Figure 3.** OWASP environment/interface.

**Figure 4.** DVWA environment.

## 4. Web Application Security Training Tutorials Used

In this section, we discuss some existing Web Application security tutorials that we have used either online or offline for analysing various mechanisms. These tutorials purposefully contain vulnerabilities for the user to discover and exploit.

**OWASP** - The Open Web Application Security Project (OWASP) is a 501c3 not-for-profit worldwide charitable organization focused on improving the security of application software [25]. Tutorials are written in Java. This tutorial covers the 10 most common Web application vulnerabilities such as (i) Injection flaws, (ii) Cross-Site Scripting (XSS), (iii) Broken Authentication and Session Management, (iv) Insecure Direct Object References, (v) Cross-Site Request Forgery (CSRF), (vi) Security Misconfiguration, (vii) Insecure Cryptographic Storage, (viii) Failure to Restrict URL Access, (ix) Insufficient Transport Layer Protection, and (x) Invalidated Redirects and Forwards. In addition, they provide hints, prevention, solution, and show java options. Every year they present the Top-10 Web Application vulnerabilities. The source code of the project and LiveCD are free of charge and accessible to almost every user. Though it provides comprehensive practices, the explanation of the topics is lacking and is left to the user to learn. It focuses more on hands-on part rather than the teaching side. Because of being completely Java-oriented, it is not concerned about applications built using other languages such as PHP or RoR (Ruby on Rails). Figure 3 shows an OWASP environment.

**DVWA** - Damn Vulnerable Web Application (**DVWA**) [26] is another practice tool built using PHP/MySQL. It is an aid for security professionals and Web developers to test and try out their skills and tools in a legal practice environment. Besides that, it is a handy approach to train/teach users (i.e., students, teachers, researchers, security professionals) on Secure Web Development. The source code and LiveCD are made available for free. This tutorial covers the following topics: Brute Force, Command Execution, CSRF, File Inclusion, SQL Injection (Blind), Upload, XSS reflected, and XSS stored. Compared to OWASP, it is less comprehensive and covers only few topics. In this tutorial, there is a lack of adequate information not only of direct topic-related discussions but also of guidelines, hints, and solutions. Users can only find information on topics through some provided Internet links/sources. Figure 4 shows the DVWA environment.

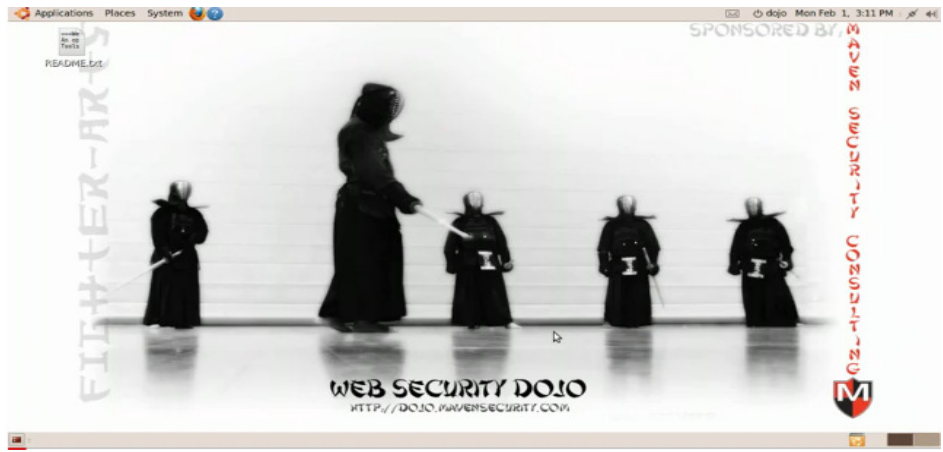

**Figure 5.** Web Security Dojo environment.

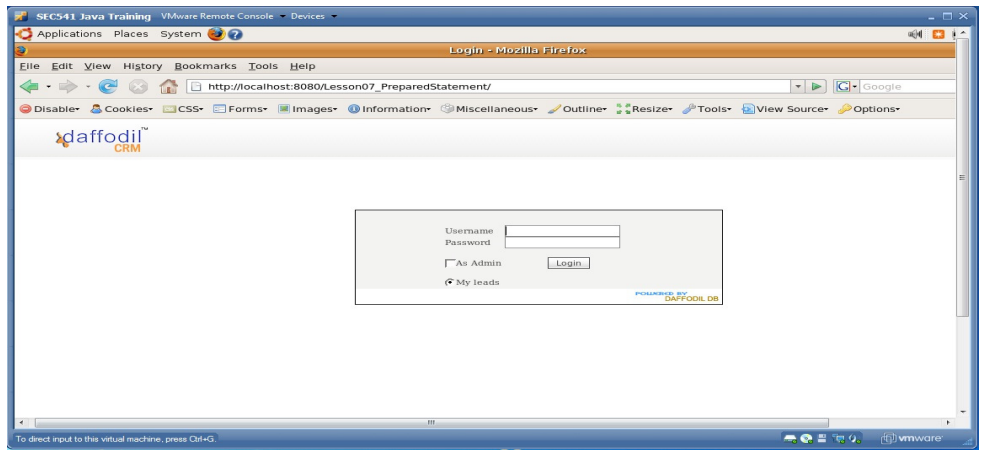

**Figure 6.** daffodil lessons.

**Web Security Dojo** - This is a "*free open-source self-contained training environment for Web Application Security penetration testing. Tools + Targets = Dojo*" [27]. The VmWare image is provided for free. Users can download and install it in a virtual machine at their own pace with full documentation. Some default targets provided are: DVWA (Damn Vulnerable Web App), REST Demos, and JSON demos. They also provide WebGoat, Hackme Casino vulnerable application, Insecure Web App and some tools like burp suite. It is a very efficient environment for practice; however, it seems to be a bit advanced tool for a security-practice beginner. The environment of Web Security Dojo is shown in Figure 5.

**Daffodil –** This is also an open source web application project designed for learning purpose [28]. It is similar to OWASP and DVWA applications. It contains both exercises and solutions for the selected Web application vulnerabilities. This tutorial also lacks proper topic discussions. The user has to look for other sources to find out more information on selected practices. It should be made more user-friendly so that beginners (i.e., the naive practitioners) can work without much hassle of finding information from here and there. Figure 6 shows a snapshot of Daffodil Lessons.

We have discussed all these training tutorials here for better understanding of the topic by the readers. In fact, for our work, we have used all these tutorials side-by-side our literature survey, problem definitions, possible solutions, analysis, and comparisons of various approaches.

## 5. Detecting SQL Injection

In order to protect a Web application from SQL Injection attacks, there are two major concerns. Firstly, there is a great need of a mechanism to detect and exactly identify SQL Injection attacks. Secondly, knowledge of SQL Injection Vulnerabilities (SQLIVs) is a must for securing a Web application. So far, many frameworks have been used and/or suggested to detect SQLIVs in Web applications. Here, we mention the prominent solutions and their working methods in brief to let the readers know about the core ideas behind each work.

### 5.1 SAFELI

Fu et al., in [3] propose a Static Analysis Framework in order to detect SQL Injection Vulnerabilities. SAFELI framework aims at identifying the SQL Injection attacks during the compile-time. This static analysis tool has two main advantages. Firstly, it does a White-box Static Analysis and secondly, it uses a Hybrid-Constraint Solver. For the White-box Static Analysis, the proposed approach considers the byte-code and deals mainly with strings. For the Hybrid-Constraint Solver, the method implements an efficient string analysis tool which is able to deal with Boolean, integer and string variables.

Table 3. SQL Injection Countermeasures.

| Countermeasure | Overview |
|---|---|
| SQL-IDS [2] | A specification based approach to detect malicious intrusions |
| Prepared Statements [4] | It is a fixed query "template" which is pre-defined, providing type–specific placeholders for input data |
| AMNESIA [6] | This scheme identifies illegal queries before their execution. Dynamically-generated queries are compared with the statically-built model using a runtime monitoring |
| SQLrand [13] | A strong random integer is inserted in the SQL keywords. |
| SQL DOM [14] | A set of classes that are strongly-typed to a database schema are used to generate SQL statements instead of string manipulation |
| SQLIA prevention using Stored Procedures [15], [16] | Combination between static analysis and runtime monitoring |
| SQLGuard [17] | The parse trees of the SQL statement before and after user input are compared at a run time. The Web script has to be modified |
| CANDID [18] | Programmer-intended query structures are guessed based upon evaluation runs over non-attacking candidate inputs |
| SQLIPA [20] | Using user name and password hash values, to improve the security of the authentication process |
| SQLCHECK [21] | A key is inserted at both beginning and end of user's input. Invalid syntactic forms are the attacks. The key strength is a major issue |
| DIWeDa [22] | To detect various types of intrusions in Web Databases applications |
| Manual approaches [23] | Defensive programming and Code review mechanisms are applied |
| Automated approaches [23] | Static analysis FindBugs and Web vulnerability scanning frameworks are implemented |

The implementation of this framework was done on ASP.NET Web applications and it was able to detect vulnerabilities that were ignored by the black-box vulnerability scanners. The methodology is an efficient approximation mechanism to deal with string constraints. However, the approach is only dedicated to ASP.NET vulnerabilities.

### 5.2 Thomas et al.'s Scheme

Thomas et al., in [4] suggest an automated prepared statement generation algorithm to remove SQL Injection Vulnerabilities (SQLIVs). They implement their research work using four open source projects namely: (i) Net-trust, (ii) ITrust, (iii) WebGoat, and (iv) Roller. Based on the experimental results, their prepared statement code was able to successfully replace 94% of the SQLIVs in four open source projects. However, the experiment was conducted using only Java with a limited number of projects. Hence, the wide application of the same approach and tool for different settings still remains an open research issue to investigate.

### 5.3 Ruse et al.'s Approach

In [5], Ruse et al. propose a technique that uses automatic test case generation to detect SQL Injection Vulnerabilities. The main idea behind this framework is based on creating a specific model that deals with SQL queries automatically. In addition, the approach identifies the relationship (dependency) between sub-queries. Based on the results, the methodology is shown to be able to specifically identify the causal set and obtain 85% and 69% reduction respectively while experimenting on few sample examples. Moreover, it does not produce any false positive or false negative and it is able to detect the real cause of the injection. In spite of the claimed and apparent efficiency of the technique, the major drawback of the work is that it was not tested with real queries on a real-life existing database.

### 5.4 Haixia and Zhihong's Database Security Testing Scheme

In [7], Haixia and Zhihong propose a secure database testing design for Web applications. They suggest a few things; firstly, detection of potential input points of SQL Injection; secondly, generation of test cases automatically, then finally finding the database vulnerability by running the test cases to make a simulation attack to an application. The proposed methodology is shown to be efficient as it was able to detect the input points of SQL Injection exactly and on time as the authors expected. However, after analyzing the scheme, we find that the approach is not a complete solution but rather it needs additional improvements in two main aspects: the detection capability and the development of the attack rule library.

### 5.5 Roichman and Gudes's Fine-grained Access Control Scheme

In [8], Roichman and Gudes, in order to secure Web application databases, suggest using a fine-grained access control to Web databases. They develop a new method based on fine-grained access control mechanism. The access to the database is supervised and monitored by the built-in database access control. This approach is efficient in the fact that the security and access control of the database is transferred from the application layer to the database layer.

Table 4. Various Schemes and SQL Injection Attacks.

| SCHEMES | Tautology | Logically Incorrect Queries | Union Query | Stored Procedure | Piggy-Backed Queries | Inference | Alternate Encodings |
|---|---|---|---|---|---|---|---|
| AMNESIA [6] | ✓ | ✓ | ✓ | x | ✓ | ✓ | ✓ |
| SQLrand [13] | ✓ | x | ✓ | x | ✓ | ✓ | x |
| SQLDOM [14] | ✓ | ✓ | ✓ | x | ✓ | ✓ | ✓ |
| WebSSARI [15,16] | ✓ | ✓ | ✓ | ✓ | ✓ | ✓ | ✓ |
| SQLGuard [17] | ✓ | ✓ | ✓ | x | ✓ | ✓ | ✓ |
| CANDID [18] | ✓ | x | x | x | x | x | x |
| SQLIPA [20] | ✓ | x | x | x | x | x | x |
| SQLCHECK [21] | ✓ | ✓ | ✓ | x | ✓ | ✓ | ✓ |
| DIWeDa [22] | x | x | x | x | x | ✓ | x |
| Automated approaches [23] | ✓ | ✓ | ✓ | x | ✓ | ✓ | x |

Table 5. Various Approaches and Types of Tasks.

| Approaches | Goals | |
|---|---|---|
|  | Detection | Prevention |
| SQL-IDS [2] | Yes | Yes |
| AMNESIA [6] | Yes | Yes |
| SQLrand [13] | Yes | Yes |
| SQL DOM [14] | Yes | Yes |
| WebSSARI [15], [16] | Yes | Yes |
| SQLGuard [17] | Yes | No |
| CANDID [18] | Yes | No |
| SQLIPA [20] | Yes | No |
| SQLCHECK [21] | Yes | No |
| DIWeDa [22] | Yes | No |

This is a solution of the vulnerability of the SQL session traceability. Besides that, it is a framework which is applicable to almost all database applications. Therefore, it significantly decreases the risk of attacks at the backend of the database application.

**5.6 Shin et al.'s Approach**

In [19], Shin et al. suggest SQLUnitGen, a Static-analysis-based tool that automate testing for identifying input manipulation vulnerabilities. They apply SQLUnitGen tool which is compared with FindBugs, a static analysis tool. The proposed mechanism is shown to be efficient (483 attack test cases) as regard to the fact that false positive was completely absent in the experiments. However for different scenarios, false negatives at a small number were noticed. In addition to that, it was found that due to some shortcomings, a more significant rate of false negatives may occur "*for other applications*". Hence, the authors talk about concentrating on getting rid of those significant false negatives and further improvement of the approach to cover input manipulation vulnerabilities as their future works.

**5.7 SQL-IDS Approach**

Kemalis and Tzouramanis in [2] suggest using a novel specification-based methodology for the detection of exploitations of SQL injection vulnerabilities. The proposed query-specific detection allowed the system to perform focused analysis at negligible computational overhead without producing false positives or false negatives. This new approach is very efficient in practice; however, it requires more experiments and comparison with available detection methods under a shared and flexible benchmarking environment.

**6. SQL Injection Countermeasures: Detection and Prevention Techniques**

In the previous section, we have discussed various schemes that only deal with SQL Injection detection. After having successfully detected any vulnerability or any kind of attack that exploits the vulnerability, other schemes could be applied to cure the system. In usual case, there are mainly two types of schemes; some are for prevention and others are for curing the system once it is under attack. In case of SQL Injection, those schemes which work for preventing SQL

Injection also do the curing of the system (or application) in early stage. Hence, in plain term, we could call the schemes '*countermeasures*'.

A strong countermeasure can remove or at least block all the available vulnerabilities in a system and thus it could protect it against various types of attacks that take advantage of the vulnerabilities. Once a system is under attack, the curing mechanisms include some other techniques like re-setting the system, re-organizing the various elements in the system, etc., which are not the topics of our current study. As those mechanisms mainly deal with other aspects of network setting, database re-shuffling, re-organizing, and utilizing clean slate approach of re-installing the system (or, application), the curing schemes are irrelevant for our survey. After our analysis of the available steps and guidelines (after-attack scenario), we found that they are more related to the managerial and administrative policies set for the system (or, application) once the attacks are launched against it and it suffers from damage.

In this section, we list a number of countermeasures that could be employed before and during running the system. It should be noted that these schemes not only detect SQL Injection but also take necessary measures so that the vulnerabilities are not exploited by the rogue entities. So, these schemes defer from the schemes mentioned in the previous section in the point that they do more than just detection of SQL Injection. Here, we also present brief descriptions and analyze each scheme from the critical '*need-to-know*' angles. Table 3 shows a summary of so far known countermeasures against SQL Injection.

Now, let us see what these schemes are actually about. The remaining texts in this section will analyze the various aspects covered in the different types of countermeasures.

### 6.1 AMNESIA

In [6], Junjin proposes AMNESIA approach for tracing SQL input flow and generating attack input, JCrasher for generating test cases, and SQLInjectionGen for identifying hotspots. The experiment was conducted on two Web applications running on MySQL1 1 v5.0.21. Based on three attempts on the two databases, SQLInjectionGen was found to give only two false negatives in one attempt. The proposed framework is efficient considering the fact that it emphasizes on attack input precision. Besides that, the attack input is properly matched with method arguments. Better than all the previous advantages, the proposed approach has no false positives and counts small number of false negatives. The only disadvantage of this approach is that it involves a number of steps using different tools.

### 6.2 SQLrand Scheme

In [13], SQLrand approach (approach using randomized SQL query language, targeting a particular *Common Gateway Interface (CGI)* application) is proposed by Boyd and Keromytis. For the implementation, they used a proof of concept proxy server in between the Web server (client) and the SQL server; they de-randomized queries received from the client and sent the request to the server. This de-randomization framework has two main advantages: portability (applied with wide range of DBMS) and security (database content highly protected). The proposed scheme has a good performance: 6.5 milliseconds is the maximum latency overhead imposed on every query. Hence, it is efficient considering the performance obtained and defense against injected queries. However, this is a proof of concept; it still requires further testing and support from programmers in building tools using SQLrand targeting more DBMS back-ends.

### 6.3 SQL DOM Scheme

SQL DOM (a set of classes that are strongly-typed to a database schema) framework is suggested by McClure and Krüger in [14]. They closely consider the existing flaws while accessing relational databases from the OOP (Object-Oriented Programming) Language's point of view. They mainly focus on identifying the obstacles in the interaction with the database via CLIs (Call Level Interfaces). SQL DOM object model is the proposed solution to tackle these issues through building a secure environment (i.e., creation of SQL statement through object manipulation) for communication. The qualitative evaluation of this approach has shown many advantages and benefits in terms of: error detection during compile time, reliability, testability, and maintainability. Though this mechanism is efficient, it can be further improved with more advanced and latest tool such as CodeSmith [31].

### 6.4 SQLIA Prevention Using Stored Procedures

Stored procedures are subroutines in the database which the applications can make call to [15]. The prevention in these stored procedures is implemented by a combination of static analysis and runtime analysis. The static analysis used for commands identification is achieved through stored procedure parser and the runtime analysis by using a SQLChecker for input identification. Huang et al. proposed in [16] a combination of static analysis and runtime monitoring to fortify the security of potential vulnerabilities. WebSSARI (Web application Security by Static Analysis and Runtime Inspection) was used and implemented on 230 open source applications on SourceForge.net. The approach was effective however it failed to remove the SQLIVs (SQL Injection Vulnerabilities). It was only able to list the input either white or black.

### 6.5 Parse Tree Validation Approach

Buehrer et al. [17] adopted the parse tree framework. They compared the parse tree of a particular statement at runtime and its original statement. They stopped the execution of statement unless there is a match. This method was tested on a student Web application using SQLGuard. Although this approach is efficient, it has two major drawbacks: additional overheard computation and listing of input only (black or white).

### 6.6 Dynamic Candidate Evaluations Approach

In [18], Bisht et al. propose CANDID (CANdidate evaluation for Discovering Intent Dynamically). It is a Dynamic Candidate Evaluations method for automatic prevention of SQL Injection attacks. This framework dynamically extracts the query structures from every SQL query location which are intended by the developer (programmer). Hence, it solves the issue of manually modifying the application to create the prepared statements. Though this tool is shown to be efficient for some cases, it fails in many other cases. For example, it is inefficient when dealing with external functions and when applied at a wrong level. Besides that, sometimes it also fails due to the limited capability of the scheme.

### 6.7 Ali et al.'s Scheme

Ali et al. [20] adopt the hash value approach to further improve the user authentication mechanism. They use the user name and password hash values. SQLIPA (SQL Injection Protector for Authentication) prototype was developed in order to test the framework. The username and password hash values are created and calculated at runtime for the first time the particular user account is created. Hash values are stored in the user account table. Though the proposed framework was tested on few sample data and had an overhead of 1.3 ms, it requires further improvement to reduce the overhead time. It also requires to be tested with larger amount of data.

### 6.8 SQLCHECKER Approach

Su and Wassermann [21] implement their algorithm with SQLCHECK on a real time environment. It checks whether the input queries conform to the expected ones defined by the programmer. A secret key is applied for the user input delimitation [1]. The analysis of SQLCHECK shows no false positives or false negatives. Also, the overhead runtime rate is very low and can be implemented directly in many other Web applications using different languages. It is a very efficient approach; however, once an attacker discovers the key, it becomes vulnerable. Furthermore, it also needs to be tested with online Web applications.

### 6.9 Detecting Intrusions in Web Databases (DIWeDa) Approach

Roichman and Gudes [22] propose IDS (Intrusion Detection Systems) for the backend databases. They use DIWeDa, a prototype which acts at the session level rather than the SQL statement or transaction stage, to detect the intrusions in Web applications. DIWeDa profiles the normal behavior of different roles in terms of the set of SQL queries issued in a session, and then compares a session with the profile to identify intrusions [22]. The proposed framework is efficient and could identify SQL injections and business logic violations too. However, with a threshold of 0.07, the True Positive Rate (TPR) was found to be 92.5% and the False Positive Rate (FPR) was 5%. Hence, there is a great need of accuracy improvement (Increase of TPR and decrease of FPR). It also needs to be tested against new types of Web attacks.

### 6.10 Manual Approaches

MeiJunjin [23] highlights the use of manual approaches in order to prevent SQLI input manipulation flaws. In manual approaches, defensive programming and code review are applied. In defensive programming: an input filter is implemented to disallow users to input malicious keywords or characters. This is achieved by using white lists or black lists. As regards to the code review [24], it is a low cost mechanism in detecting bugs; however, it requires deep knowledge on SQLIAs.

### 6.11 Automated Approaches

Besides using manual approaches, MeiJunjin [23] also highlights the use of automated approaches. The author notes that the two main schemes are: Static analysis FindBugs and Web vulnerability scanning. Static analysis FindBugs approach detects bugs on SQLIAs, gives warning when an SQL query is made of variable. However, for the Web vulnerability scanning, it uses software agents to crawl, scans Web applications, and detects the vulnerabilities by observing their behavior to the attacks.

### 6.12 Comparisons

It would be difficult to give a clear verdict which scheme or approach is the best as each one has some proven benefits for specific types of settings (i.e., systems). Hence, in this section, we note down how various schemes work against the identified SQL Injection attacks. Table 4 shows a chart of the schemes and their defense capabilities against various SQLIAs. This table shows the comparative analysis of the SQL Injections prevention techniques and the attack types. Though many approaches have been identified as detection or prevention techniques, only few of them were implemented in practicality. Hence, this comparison is not based on empirical experience but rather it is an analytical evaluation.

In Table 5, we note down the major approaches to deal with SQL Injection and classify them based on their features.

## 7. Conclusion and Future Research Directions

Though many approaches and frameworks have been identified and implemented in many interactive Web applications, security still remains a major issue. SQL Injection prevails as one of the top-10 vulnerabilities and threat to online businesses targeting the backend databases. In this paper, we have reviewed the most popular existing SQL Injections related issues.

Key findings of this study could be summarized as:
- Detailed survey report on various types of SQL Injection attacks, vulnerabilities, detection, and prevention techniques
- Assessment of techniques based on their performance and practicality
- Awareness information of the threat of SQL Injections by providing recent and updated cases and information
- Exploration of "Web Application Security training tutorials" to train security practitioners to deal with SQL Injection attacks

The findings of this study could be used for penetration testing purpose in order to protect data either in academic or industrial fields. Our research outcomes help:
- to measure the security level of Web Applications using proposed tools
- to find/detect vulnerabilities of online applications
- to protect applications against using proposed secure coding approaches
- to train security practitioners on SQL Injection using the proposed tutorials

We believe that the work would be useful both for the general readers of the topic as well as for the practitioners. As a future work, we would like to develop a countermeasure that can efficiently tackle the innovative SQL Injection attacks and fix as much vulnerability as possible. Hackers are in reality very innovative and as the time is passing by, new attacks are being launched that may need new ways of thinking about the solutions we currently have at our hands.

## Acknowledgement

This project was supported by NDC Lab, KICT, IIUM.